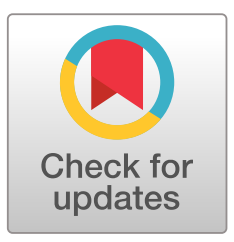

ARTICLE

# Robust Backstepping Control of a Quadrotor Unmanned Aerial Vehicle under Colored Noises

**Mehmet Karahan***

Electrical and Electronic Engineering Department, Atilim University, Ankara, 06830, Turkey

*Corresponding Author: Mehmet Karahan. Email: mehmet.karahan@atilim.edu.tr

**ABSTRACT**

Advances in software and hardware technologies have facilitated the production of quadrotor unmanned aerial vehicles (UAVs). Nowadays, people actively use quadrotor UAVs in essential missions such as search and rescue, counter-terrorism, firefighting, surveillance, and cargo transportation. While performing these tasks, quadrotors must operate in noisy environments. Therefore, a robust controller design that can control the altitude and attitude of the quadrotor in noisy environments is of great importance. Many researchers have focused only on white Gaussian noise in their studies, whereas researchers need to consider the effects of all colored noises during the operation of the quadrotor. This study aims to design a robust controller that is resistant to all colored noises. Firstly, a nonlinear quadrotor model was created with MATLAB. Then, a backstepping controller resistant to colored noises was designed. The designed backstepping controller was tested under Gaussian white, pink, brown, blue, and purple noises. PID and Lyapunov-based controller designs were also carried out, and their time responses (rise time, overshoot, settling time) were compared with those of the backstepping controller. In the simulations, time was in seconds, altitude was in meters, and roll, pitch, and yaw references were in radians. Rise and settling time values were in seconds, and overshoot value was in percent. When the obtained values are examined, simulations prove that the proposed backstepping controller has the least overshoot and the shortest settling time under all noise types.

## 1 Introduction

Quadrotor is a four-rotor UAV with vertical take-off and landing (VTOL) capability [1,2]. Quadrotor does not need a runway during take-off and landing. It can hang in the air, rotate around its own axis, flip in the air, and achieve hard maneuvers [3,4]. These characteristics make quadrotor UAVs more advantageous than fixed-wing UAVs [5].

Today, quadrotors are used in various areas, such as search and rescue, surveillance, counter-terrorism, border patrol, combating natural disasters, agricultural spraying, cargo transportation, mapping, mineral exploration, and aerial photography [6,7].

In order for the quadrotor to perform these critical tasks, it needs a robust controller that can operate in noisy environments [8,9]. Many researchers have carried out studies on controller design for quadrotor UAVs that can operate in noisy environments. However, since researchers generally focus on the quadrotor's trajectory tracking under white Gaussian noise in their studies, there have been few studies in the literature on the quadrotor's trajectory tracking under other colors in the color spectrum. For a comprehensive robustness analysis, it is necessary to consider all colored noises. Fig. 1 shows the colors of noise.

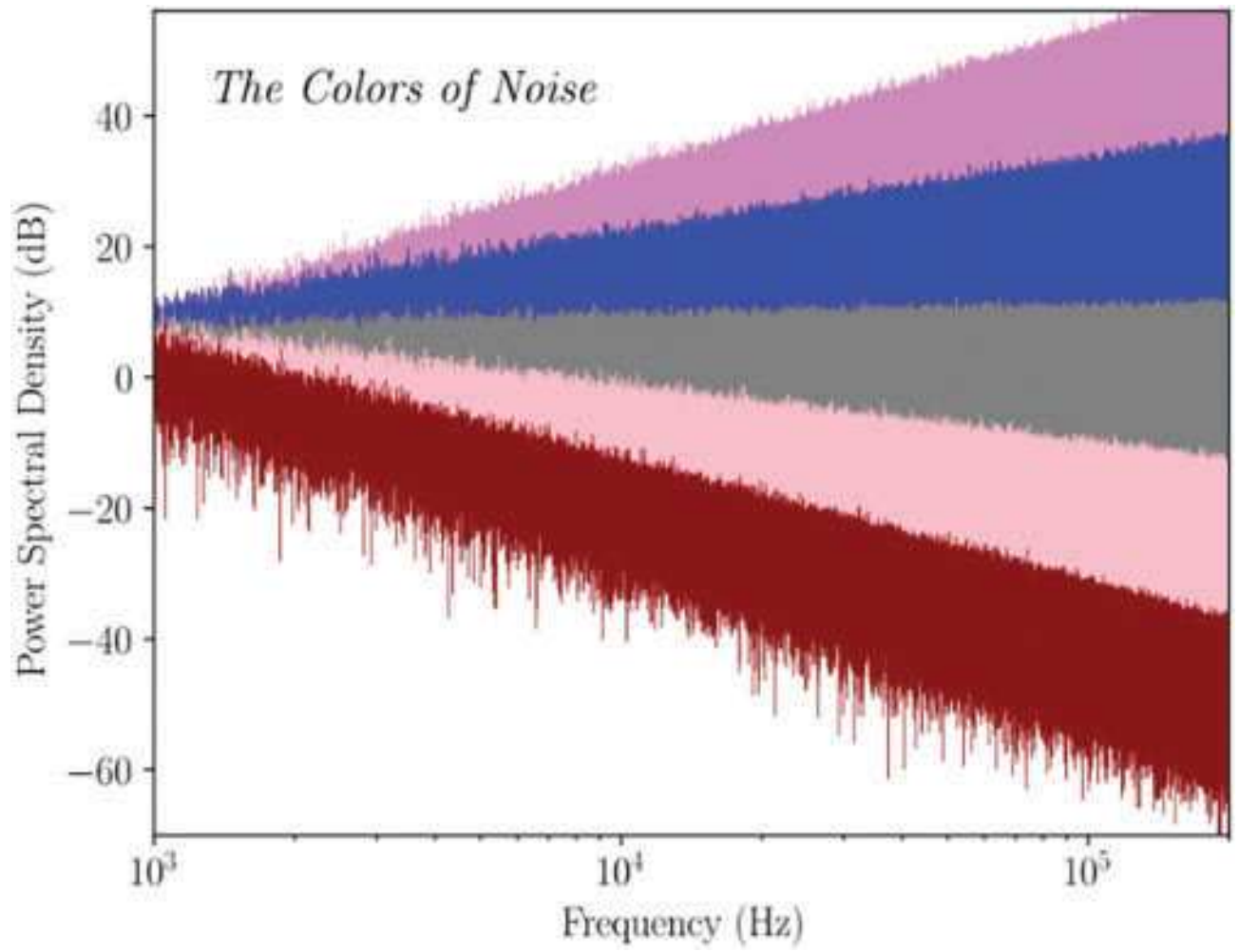

**Figure 1:** Power spectral densities of different colors

The primary motivation of this research is to design a robust controller for a quadrotor UAV that can track altitude and attitude references under all colored noises. The main contributions of this study are summarized below:

- Firstly, a nonlinear model of the quadrotor UAV was created with MATLAB/Simulink software.
- Secondly, a robust backstepping controller was designed to control the altitude and attitude of this nonlinear quadrotor model. A classical PID controller design and a Lyapunov-based controller design were also used to perform a comparative robustness analysis with the backstepping controller.
- Thirdly, these controllers were tested under band-limited Gaussian white, pink, brown, blue, and purple noises. The controllers' overshoot, rise time, and settling time values under these noises were acquired, and a comparative robustness analysis was performed.

The obtained results demonstrated the robustness of the backstepping controller. A Pseudocode briefly explaining the work done is shown in Fig. 2.

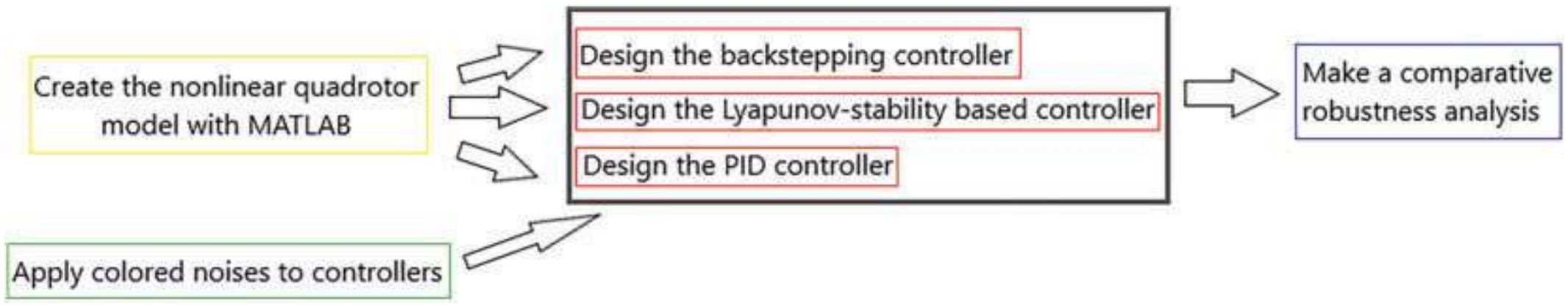

**Figure 2:** Pseudocode of the work

## 2 Related Work

Researchers have worked on different techniques to control the altitude and attitude of the quadrotor. This section discusses the recent related studies on controlling the quadrotor. MA et al. focused on particle swarm optimization tuned trajectory tracking control for a quadrotor based on the Lyapunov method under the Gaussian white noise effect [10]. Kumar et al. used the Lyapunov stability analysis and developed a novel dynamic recurrent functional link neural network for the identification of nonlinear systems [11]. Kumar et al. conducted a comparative study of neural networks for the control of nonlinear dynamical systems using Lyapunov stability-based adaptive learning rates [12]. Cheded et al. developed a novel, robust Kalman filter-based controller for the trajectory tracking of a quadrotor under a Gaussian zero mean white noise [13]. Zhao et al. focused on the reference tracking of a quadrotor under white noise. They have developed an active disturbance rejection switching control algorithm against white noise. Their algorithm could reach stable control of the quadrotor UAV in the white noise range of 0.1 dB [14]. Guerrero-Sánchez et al. developed an observer-based controller for a quadrotor. Their designed observer estimates the state from noisy output measurements. They used a Gaussian noise with a variance of 0.001 as the test signal [15]. Mahfouz et al. focused on the control of a four-rotor UAV under white Gaussian noise. They compared the performances of different PID control techniques under white Gaussian noise [16]. Wang et al. designed a PD controller to control the attitude of the quadrotor. They used a white Gaussian noise with zero mean and one variance to test the robustness of the controller [17]. Hou et al. proposed a nonsingular terminal sliding mode controller for a quadrotor and tested their proposed sliding mode controller under Gaussian white noise and model uncertainties [18]. Xu et al. focused on the path-following control of a quadrotor under Gaussian noise. They used the Lyapunov theorem while designing the controller [19]. Cen et al. proposed a Gaussian process regression-based control technique for a quadrotor UAV. They performed trajectory tracking simulations of the quadrotor under measurement noise and compared their proposed controller with an integral backstepping controller [20]. Labbadi et al. designed a sliding mode controller for a quadrotor under Gaussian random disturbances and uncertainties. They ensured the stability of the quadrotor by using Lyapunov's theorem [21]. Arul et al. presented a collision avoidance algorithm and utilized a model predictive control technique for quadrotor swarms under Gaussian noise. Their proposed method requires 5 ms to compute a local collision-free trajectory [22]. Bouaiss et al. designed a Model Predictive Controller (MPC) for a quadrotor. They used the MPC controller for trajectory tracking under Gaussian noise and compared the performance of the MPC controller with a classic PID controller [23]. Wang et al. developed an MPC controller for a quadrotor's trajectory planning problem in unknown environments. They implemented a real-time Gaussian noise as an aerodynamic disturbance to the quadrotor. They achieved a trajectory generation in unknown places by up to 75% in their tests [24]. Noordin et al. designed a super-twisting sliding mode controller for a quadrotor. They tested the robustness of the controller under Gaussian white noise. They compared the trajectory simulation results of the proposed sliding mode controller and classical PID controller. They saw that the sliding mode controller provided more robust performance than the PID controller [25].

## 3 Nonlinear Modelling of the Quadrotor

A quadrotor is a UAV with four propellers that can take off and land vertically, hover in the air, and rotate around its own axis [26,27].

In this research, a quadrotor with a cross configuration was used [28]. The schematic representation of the quadrotor used is given in Fig. 3. Rotors rotating in the same direction are shown with the same color. The body axis, the Earth axis, the direction of the torques produced, and the attitude angles are shown in Fig. 3.

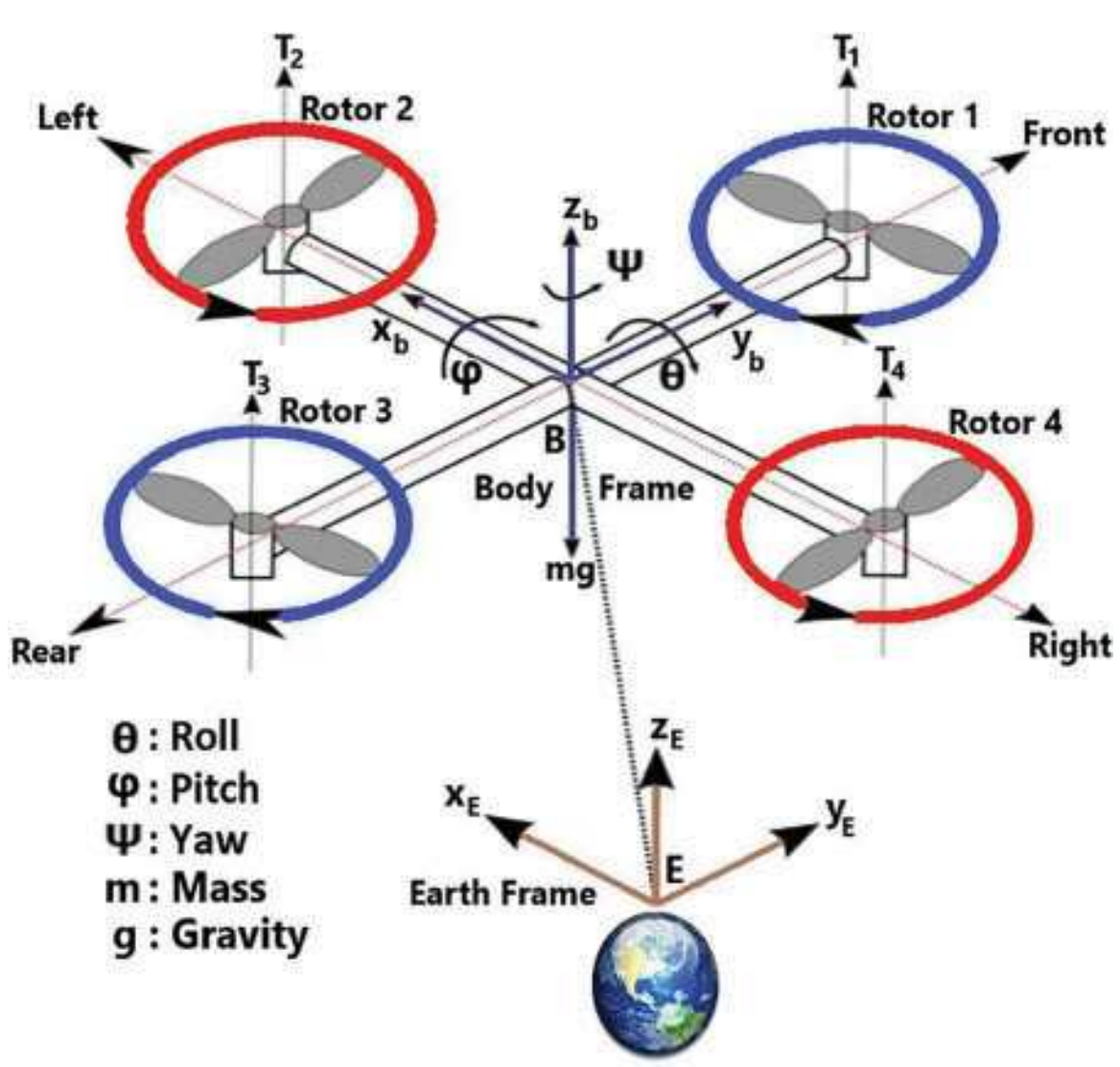

**Figure 3:** Schematic representation of the quadrotor

The quadrotor moves in 6 degrees of freedom [29]. The rotation matrix is used in the transformation from the Earth-centered frame to the body-centered frame. Rotation matrices for attitude angles are given in Eqs. (1) to (4). In the rotation matrices below, $c$ stands for cosine, and $s$ stands for sine.

$$R(\varphi) = \begin{bmatrix} 1 & 0 & 0 \\ 0 & c(\varphi) & s(\varphi) \\ 0 & -s(\varphi) & c(\varphi) \end{bmatrix} \tag{1}$$

$$R(\theta) = \begin{bmatrix} c(\theta) & 0 & -s(\theta) \\ 0 & 1 & 0 \\ s(\theta) & 0 & c(\theta) \end{bmatrix} \tag{2}$$

$$R(\psi) = \begin{bmatrix} c(\psi) & s(\psi) & 0 \\ -s(\psi) & c(\psi) & 0 \\ 0 & 0 & 1 \end{bmatrix} \tag{3}$$

$$R(\varphi, \theta, \psi) = R(\varphi)\, R(\theta)\, R(\psi) \tag{4}$$

Eq. (5) shows the orthogonal rotation matrix.

$$R = \begin{bmatrix} c\psi c\theta & s\psi c\theta & -s\theta \\ c\psi s\theta s\varphi - s\psi c\varphi & s\psi s\theta s\varphi + c\psi c\varphi & c\theta s\varphi \\ c\psi s\theta c\varphi + s\psi s\varphi & s\psi s\theta c\varphi - c\psi s\varphi & c\theta c\varphi \end{bmatrix} \tag{5}$$

Eqs. (6) and (7) represent the transformation between body angle rates and Earth angular rates. $P$, $Q$, and $R$ are angular velocities in the body frame. $\dot{\varphi}$, $\dot{\theta}$, and $\dot{\psi}$ represent Earth's angular rates.

$$T = \begin{bmatrix} 1 & \tan(\theta)\, s(\varphi) & \tan(\theta)\, c(\varphi) \\ 0 & c(\varphi) & -s(\varphi) \\ 0 & \sec(\theta)\, s(\varphi) & \sec(\theta)\, c(\varphi) \end{bmatrix} \tag{6}$$

$$\begin{bmatrix} \dot{\varphi} \\ \dot{\theta} \\ \dot{\psi} \end{bmatrix} = T \begin{bmatrix} p \\ q \\ r \end{bmatrix} \tag{7}$$

Eq. (7) requires that $\theta \neq \frac{\pi}{2}$ for derivatives. While $\varphi$ and $\theta$ angles are close to 0, it means that the quadrotor is hovering and $T$ is approximately a unit matrix [30]. In this situation, the relation between Earth's angular rates and body angular velocities can be described as linear, as in Eq. (8).

$$\begin{bmatrix} \dot{\varphi} \\ \dot{\theta} \\ \dot{\psi} \end{bmatrix} \approx \begin{bmatrix} p \\ q \\ r \end{bmatrix} \tag{8}$$

Force (F) and torque (T) formulas are represented in Eqs. (9) and (10). w is angular velocity, $b$ is thrust coefficient, and $d$ is drag coefficient. The relative speed of the rotor ($w_r$) is given in Eq. (11). The i subscript in the equations describes the number of rotors from 1 to 4.

$$F_i = bw_i^2 \tag{9}$$

$$T_i = dw_i^2 \tag{10}$$

$$w_r = -w_1 + w_2 - w_3 + w_4 \tag{11}$$

The relation between control inputs ($U_1$, $U_2$, $U_3$, $U_4$) and angular velocities is represented in Eq. (12). $U_1$ is lift force, and $U_2$, $U_3$, and $U_4$ represent relevant torques. The $l$ symbol is the arm length of the quadrotor.

$$u = \begin{bmatrix} U_1 \\ U_2 \\ U_3 \\ U_4 \end{bmatrix} = \begin{bmatrix} F \\ T_\varphi \\ T_\theta \\ T_\psi \end{bmatrix} = \begin{bmatrix} b & b & b & b \\ 0 & -lb & 0 & lb \\ lb & 0 & -lb & 0 \\ -d & d & -d & d \end{bmatrix} \begin{bmatrix} w_1^2 \\ w_2^2 \\ w_3^2 \\ w_4^2 \end{bmatrix} \tag{12}$$

The transformation from angular velocities to control inputs is given in Eq. (13).

$$\begin{bmatrix} w_1^2 \\ w_2^2 \\ w_3^2 \\ w_4^2 \end{bmatrix} = \begin{bmatrix} 1/4b & 0 & 1/2bl & -1/4d \\ 1/4b & -1/2bl & 0 & 1/4d \\ 1/4b & 0 & -1/2bl & -1/4d \\ 1/4b & 1/2bl & 0 & 1/4d \end{bmatrix} \begin{bmatrix} U_1 \\ U_2 \\ U_3 \\ U_4 \end{bmatrix} \tag{13}$$

The inertial moments of the quadrotor are represented in Eqs. (14)–(16). $M_{sphere}$ is the mass of the spherical dense center, $M_{rotor}$ is the mass of one rotor, and $r$ is the radius.

$$I_x = \frac{2}{5} M_{sphere} r^2 + 2l^2 M_{rotor} \tag{14}$$

$$I_y = \frac{2}{5} M_{sphere} r^2 + 2l^2 M_{rotor} \tag{15}$$

$$I_z = \frac{2}{5} M_{sphere} r^2 + 4l^2 M_{rotor} \tag{16}$$

Fig. 4 represents the spherical mass and point masses of the quadrotor.

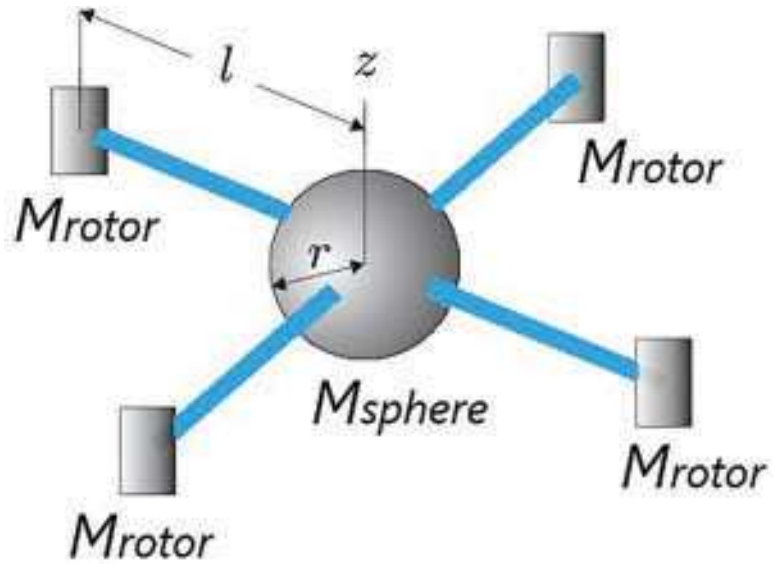

**Figure 4:** Spherical mass and point masses of the quadrotor

Equations of the motion of the quadrotor are presented from Eqs. (17) to (22).

$$\ddot{X} = \frac{U_1}{m} \left[ c(\varphi)\, s(\theta)\, c(\psi) + s(\varphi)\, s(\psi) \right] \tag{17}$$

$$\ddot{Y} = \frac{U_1}{m} \left[ s(\theta)\, s(\psi)\, c(\varphi) - s(\varphi)\, c(\psi) \right] \tag{18}$$

$$\ddot{Z} = -g + \frac{U_1}{m} \left[ c(\varphi)\, c(\theta) \right] \tag{19}$$

$$\ddot{\varphi} = \frac{1}{I_x} U_2 + \left( \frac{I_y - I_z}{I_x} \right) \dot{\psi}\dot{\theta} + w_r \frac{J_R}{I_x} \dot{\theta} \tag{20}$$

$$\ddot{\theta} = \frac{1}{I_y} U_3 + \left( \frac{I_z - I_x}{I_y} \right) \dot{\theta}\dot{\psi} + w_r \frac{J_R}{I_y} \dot{\varphi} \tag{21}$$

$$\ddot{\psi} = \frac{1}{I_z} U_4 + \left( \frac{I_x - I_y}{I_z} \right) \dot{\theta}\dot{\varphi} \tag{22}$$

In this study, the OS4 quadrotor model is used. Table 1 shows the physical constants of this quadrotor.

**Table 1:** Physical constants of the OS4 quadrotor model

| Constant | Value | Unit |
|---|---|---|
| Arm length (l) | 0.23 | m |
| Gravity (g) | 9.81 | $m/s^2$ |
| Mass (m) | 0.65 | kg |
| Max. rotor velocity ($w_{max}$) | 1000 | rad/sec |
| Max. torque ($t_{max}$) | 0.15 | Nm |
| Thrust coefficient (b) | 3.13 | $Ns^2$ |
| Drag coefficient (d) | $7.5 \times 10^{-7}$ | $Ns^2$ |
| Inertial moment on $x$ axis ($I_x$) | $7.5 \times 10^{-3}$ | $kg.m^2$ |
| Inertial moment on $y$ axis ($I_y$) | $7.5 \times 10^{-3}$ | $kg.m^2$ |
| Inertial moment on $z$ axis ($I_z$) | $1.3 \times 10^{-2}$ | $kg.m^2$ |
| Rotor inertia ($J_R$) | $6.5 \times 10^{-5}$ | $kg.m^2$ |

## 4 Control Structures

In this part, PID, Lyapunov-based, and backstepping controller structures are described.

### 4.1 PID Control

PID controllers are widely used in the industry because their structure is simple, their parameters can be easily adjusted, and they are easy to manufacture [31]. To prove the robustness of the proposed backstepping controller in this research, a comparative robustness analysis was carried out with the PID controller. The PID controller has three parameters: $K_p$, $K_i$, and $K_d$. The PID controller aims to minimize the error signal $e(t)$ by using the control signal $u(t)$. The general equation of the PID controller is given in Eq. (23).

$$u(t) = K_p e(t) + K_i \int_0^t e(\tau)\, d\tau + K_d \frac{de(t)}{dt} \tag{23}$$

The $U_1$ control input that controls the altitude of the quadrotor is given in Eq. (24).

$$U_1 = \frac{m\left(g + e_z K_p + K_i \int e_z dt + K_d \frac{de_z}{dt}\right)}{\cos\varphi \cos\theta} \tag{24}$$

In Eq. (24), $\theta$ angle is rotation around the $y$-axis and $\varphi$ angle corresponds to rotation around the $x$-axis. Since the quadrotor does not rotate in the $x$ and $y$ axes during vertical flight, the denominator of the $U_1$ will never be zero.

$U_2$, $U_3$, and $U_1$ control inputs control roll, pitch, and yaw angles, respectively. The equations of these control inputs are given in Eqs. (25)–(27).

$$U_2 = e_\varphi K_p + K_i \int e_\varphi dt + K_d \frac{de_\varphi}{dt} \tag{25}$$

$$U_3 = e_\theta K_p + K_i \int e_\theta dt + K_d \frac{de_\theta}{dt} \tag{26}$$

$$U_4 = e_\psi K_p + K_i \int e_\psi dt + K_d \frac{de_\psi}{dt} \tag{27}$$

Table 2 shows the PID controller coefficients. The PID Tuner app in Simulink was used to determine the coefficients of the PID controller. Coefficients were manually adjusted using the graphical interface of PID Tuner. Optimum controller coefficients were set to have low overshoot and short settling time.

**Table 2:** PID controller coefficients

| Coefficient | Altitude | Roll | Pitch | Yaw |
|---|---|---|---|---|
| $K_p$ | 0.82 | 0.12 | 0.14 | 0.13 |
| $K_i$ | 1 | 0.05 | 0.07 | 0.05 |
| $K_d$ | 1.65 | 0.06 | 0.08 | 0.1 |

### *4.2 Lyapunov-Based Control*

Lyapunov-based control is a nonlinear control method that depends on the Lyapunov stability theorem [32]. In this study, the most commonly used sum of squares function was used as the Lyapunov function. The Lyapunov-based controller targets to control the position of the quadrotor directly. In this approach, $x = 0$ is defined as the equilibrium point. $D$ is determined as a compact space of $f(0)$ in $R^n$ real coordinate space. The continuous Lyapunov function $V: D \rightarrow R^+$, which satisfies requirements in Eqs. (28) and (29), is defined.

$$V(0) = 0,\ V(x) > 0 \ in\, D,\ x \neq 0 \tag{28}$$

$$\dot{V}(x) \leq 0 \ in\, D \tag{29}$$

The equilibrium point is asymptotically stable in $D$ under $\dot{V}(x) < 0$ in $D$, $x \neq 0$ conditions. Then, a part that contains stabilization angles and their derivates is defined as the desired attitude at the equilibrium point for the attitude control. In this state, $x = (\varphi_d, 0, \theta_d, 0, \psi_d, 0)$, where $\varphi_d$, $\theta_d$, and $\psi_d$, are defined as desired roll, pitch, and yaw angles. Since angular velocities will be zero at the stabilization point, their derivatives will also be zero. Positive defined Lyapunov function at the desired attitude is described in Eq. (30).

$$V(x) = \frac{1}{2}\left[\dot{\varphi}^2 + (\varphi - \varphi_d)^2 + \dot{\theta}^2 + (\theta - \theta_d)^2 + \dot{\psi}^2 + (\psi - \psi_d)^2\right] \tag{30}$$

Eq. (31) describes the derivative of $V(x)$.

$$\dot{V}(x) = \left[(\varphi - \varphi_d)\,\dot{\varphi} + \ddot{\varphi}\dot{\varphi} + (\theta - \theta_d)\,\dot{\theta} + \ddot{\theta}\dot{\theta} + (\psi - \psi_d)\,\dot{\psi} + \ddot{\psi}\dot{\psi}\right] \tag{31}$$

Equations of motion defined in Eqs. (17–22) can be simplified under perfect cross configuration VTOL ($I_x = I_y$) condition when the quadrotor close to the equilibrium point ($w_r = 0$, $\dot{\psi} = 0$, $\dot{\theta} = 0$, $\dot{\varphi} = 0$) and Eq. (32) is obtained.

$$\dot{V}(x) = (\varphi - \varphi_d)\dot{\varphi} + \dot{\varphi}\frac{l}{I_x}U_2 + (\theta - \theta_d)\dot{\theta} + \dot{\theta}\frac{l}{I_y}U_3 + (\psi - \psi_d)\dot{\psi} + \dot{\psi}\frac{l}{I_z}U_4 \tag{32}$$

The control inputs of attitude angles are defined from Eqs. (33) to (35) for the stability criteria.

$$U_2 = -\frac{I_x}{l}(\varphi - \varphi_d) - k_1\dot{\varphi} \tag{33}$$

$$U_3 = -\frac{I_y}{l}(\theta - \theta_d) - k_2\dot{\theta} \tag{34}$$

$$U_4 = -I_z(\psi - \psi_d) - k_3\dot{\psi} \tag{35}$$

By substituting $U_2$, $U_3$, and $U_1$ control inputs in Eq. (32), the equation could be rewritten as in Eq. (36).

$$\dot{V}(x) = -\dot{\varphi}^2\frac{l}{I_x}k_1 - \dot{\theta}^2\frac{l}{I_y}k_2 - \dot{\psi}^2\frac{l}{I_z}k_3 \tag{36}$$

In the above equations from Eqs. (33) to (36), $k_1$, $k_2$ and $k_3$ indicate positive coefficients which are negative semidefinite. Stability for the equilibrium point is provided by the Lyapunov theorem. Asymptotic stability is achieved by LaSalle's invariance principle since the controlled maximum invariant set of the subsystem in $\mathrm{S} = \{\mathrm{X} \in \Re^6\, \dot{V}|_{\mathrm{X}} = 0\}$ is limited by the equilibrium point. Lyapunov function and its time derivative are given in Eqs. (37) and (38) for altitude control.

$$V(x) = \frac{1}{2}\left[(z - z_d)^2 + \dot{z}^2\right] \tag{37}$$

$$\dot{V}(x) = (z - z_d)\dot{z} + \dot{z}\left(g - (\cos\theta\cos\varphi)\frac{U_1}{m}\right) \tag{38}$$

$\mathrm{U}_1$ altitude control input is defined in Eq. (39) for stability.

$$U_1 = -\frac{m}{\cos\theta\cos\varphi}(z_d - z - g) - k_z\dot{z} \tag{39}$$

After the $U_1$ is substituted in Eq. (38), the Eq. (40) is obtained. $k_z$ is a positive constant given by Eq. (39), which is negative semi-definite.

$$\dot{V}(x) = -\dot{z}^2\frac{k_z}{m}(\cos\theta\cos\varphi) \tag{40}$$

The coefficients of the Lyapunov-based controller are given in Table 3. While $k_z$ is the altitude controller coefficient, $k_1$, $k_2$, and $k_3$ are the roll, pitch, and yaw angle controller coefficients. The Lyapunov-based controller coefficients are determined by trial and error. While determining the coefficients, the coefficients that would cause the least overshoot and the shortest settling time were preferred.

**Table 3:** Coefficients of the Lyapunov-based controller

| Coefficient | Value |
|---|---|
| $k_z$ | 2.15 |
| $k_1$ | 0.167 |
| $k_2$ | 0.168 |
| $k_3$ | 0.104 |

### *4.3 Backstepping Control*

The proposed backstepping control method is an adaptive control approach used in nonlinear systems. This control method depends on a recursive design that links the selection of a Lyapunov function with a feedback control system and provides strict feedback to obtain asymptotic stability. In this research, Lyapunov's direct method is combined with principles of adaptive control. First, the tracking error $z_1$ is defined in Eq. (41).

$$z_1 = \varphi_d - \varphi \tag{41}$$

The Lyapunov function and its time derivative for the $z_1$ variable are given in Eqs. (42) and (43).

$$V(z_1) = \frac{1}{2}{z_1}^2 \tag{42}$$

$$\dot{V}(z_1) = z_1(\dot{\varphi}_d - \dot{\varphi}) \tag{43}$$

Since the derivative of the Lyapunov function should be negative semi-definite, the new virtual control input $\dot{\varphi}$ is defined to stabilize $z_1$ as in Eq. (44).

$$\dot{\varphi} = \dot{\varphi}_d + a_1 z_1 \tag{44}$$

$a_1$ should be a positive coefficient to provide negative semi-definitiveness. When this virtual control input is substituted in Eq. (43), the equation in Eq. (45) is obtained.

$$\dot{V}(z_1) = -a_1{z_1}^2 \tag{45}$$

Another change of variable is represented in Eq. (46).

$$z_2 = \dot{\varphi} - \dot{\phi}_d - a_1 z_1 \tag{46}$$

After the changes, the augmented Lyapunov function can be written as in Eq. (47).

$$V(z_1, z_2) = \frac{1}{2}{z_1}^2 + \frac{1}{2}{z_2}^2 \tag{47}$$

Derivative of the Lyapunov function in Eq. (47) could be written as in Eq. (48).

$$\dot{V}(z_1, z_2) = -a_1{z_1}^2 - z_1 z_2 + z_2\ddot{\varphi} - z_2(\ddot{\varphi}_d - a_1(z_2 + a_1 z_1)) \tag{48}$$

According to Eq. (20), the $\ddot{\varphi}$ variable can be rewritten in Eq. (49).

$$\ddot{\varphi} = \dot{\psi}\dot{\theta}a_1 + a_2\dot{\theta}w_r + \frac{l}{I_x}U_2 \tag{49}$$

$U_2$ control input is defined as in Eq. (50) under $\ddot{\varphi} = 0$, $\dddot{\varphi} = 0$, $\ddot{\theta}_d = 0$, and $\dot{V}(z_1, z_2) < 0$ conditions.

$$U_2 = \frac{I_x}{l}\left(z_1 - a_1\dot{\theta}\dot{\psi} - a_2\dot{\theta}w_r - a_1\left(z_2 + a_1z_1\right) - a_2z_2\right) \tag{50}$$

The $a_2z_2$ term under $a_2 > 0$ condition is added to stabilize $z_1$. With the same steps, $U_3$ pitch angle control input and $U_4$ yaw angle control input are defined in Eqs. (51) and (52):

$$U_3 = \frac{I_y}{l}\left(z_3 - a_3\dot{\varphi}\dot{\psi} - a_4\dot{\varphi}w_r - a_3\left(z_4 + a_3z_3\right) - a_4z_4\right) \tag{51}$$

$$U_4 = I_z\left(z_5 - a_5\dot{\varphi}\dot{\theta} - a_5\left(z_6 + a_5z_5\right) - a_6z_6\right) \tag{52}$$

The Eqs. (53) to (56) describe the variables used in $U_3$ and $U_4$.

$$z_3 = \theta_d - \theta \tag{53}$$

$$z_4 = \dot{\theta} - \dot{\theta}_d - a_3z_3 \tag{54}$$

$$z_5 = \psi_d - \psi \tag{55}$$

$$z_6 = \dot{\psi} - \dot{\psi}_d - a_5z_5 \tag{56}$$

The $z_7$ variable in Eq. (57) describes the altitude tracking error.

$$z_7 = z - z_d \tag{57}$$

The Lyapunov function and its derivative for $z_7$ variable are given in Eqs. (58) and (59).

$$V\left(z_7\right) = \frac{1}{2}{z_7}^2 \tag{58}$$

$$\dot{V}\left(z_7\right) = z_7\left(\dot{z}_d - \dot{z}\right) \tag{59}$$

The $x_8$ virtual control input is defined as in Eq. (60) to stabilize $z_7$ function.

$$x_8 = \dot{z}_d + a_7z_7 \tag{60}$$

Another variable change is given in Eq. (61).

$$z_8 = x_8 - \dot{z}_d - a_7z_7 \tag{61}$$

After the variable changes, the new Lyapunov function could be written as in Eq. (62).

$$V\left(z_7, z_8\right) = \frac{1}{2}{z_7}^2 + \frac{1}{2}{z_8}^2 \tag{62}$$

The derivative of the Lyapunov function in Eq. (62) is explained as in Eq. (63).

$$\dot{V}\left(z_7, z_8\right) = -a_7{z_7}^2 - z_7z_8 + z_8x_8 - z_8\left(\ddot{z}_d - a_7\left(z_8 + a_7z_7\right)\right) \tag{63}$$

The derivative of the virtual control input $x_8$ is explained in Eq. (64).

$$\dot{x}_8 = g - \cos\theta\cos\varphi\frac{U_1}{m} \tag{64}$$

The $U_1$ altitude control input is given in Eq. (65).

$$U_1 = \frac{m}{\cos\theta\cos\varphi}\left(z_7 + g - a_7\left(z_8 + a_7 z_7\right) - a_8 z_8\right) \tag{65}$$

Table 4 represents the coefficients of the backstepping controller. The coefficients of the backstepping controller were obtained by trial and error. The aim was to get the coefficients with minimum overshoot and the fastest settling time.

**Table 4:** Backstepping controller coefficients

| Variable | Roll | Pitch | Yaw | Altitude |
|---|---|---|---|---|
| ($a_1$, $a_2$, $a_3$, $a_4$, $a_5$, $a_6$, $a_7$, $a_8$) | (8.6, 6.9) | (8.1, 3.9) | (8.4, 4.1) | (1.4, 5.9) |

## 5 Simulations

In this section, reference tracking simulations were performed under white Gaussian noise, pink noise, brown noise, blue noise, and purple noise. Simulations were made using the MATLAB program. The block diagram of the designed system is given in Fig. 5. $\varphi_d$, $\theta_d$, $\psi_d$ are reference trajectories, and $\varphi_d$, $\theta_d$, $\psi_d$ are output trajectories of the system. Colored noise is added to the output of the backstepping controller. Within the scope of the study, the proposed backstepping controller was compared with the classical PID controller and the Lyapunov-based controller. For this purpose, the rise time, overshoot, and settling time data of all three controllers were compared. The obtained results prove the robustness of the backstepping controller.

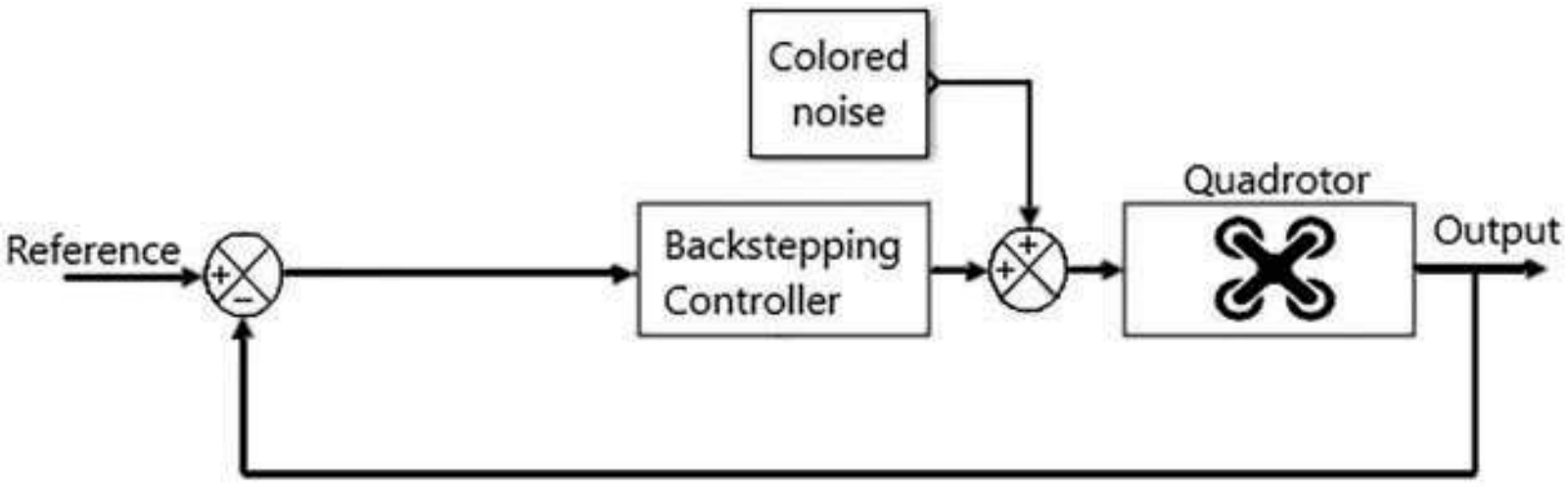

**Figure 5:** Block diagram of the designed system

### *5.1 Simulations under White Gaussian Noise*

In this section, reference tracking simulations were performed under band-limited white Gaussian noise. White Gaussian noise was given to the system with 0.01 power and 0.1 sampling time. Comparative robustness analysis was performed by obtaining rise time, overshoot and settling time data of PID, Lyapunov-based and backstepping controllers. Fig. 6 shows reference tracking simulations under white Gaussian noise.

Table 5 shows the time response data of the controllers under band-limited white Gaussian noise.

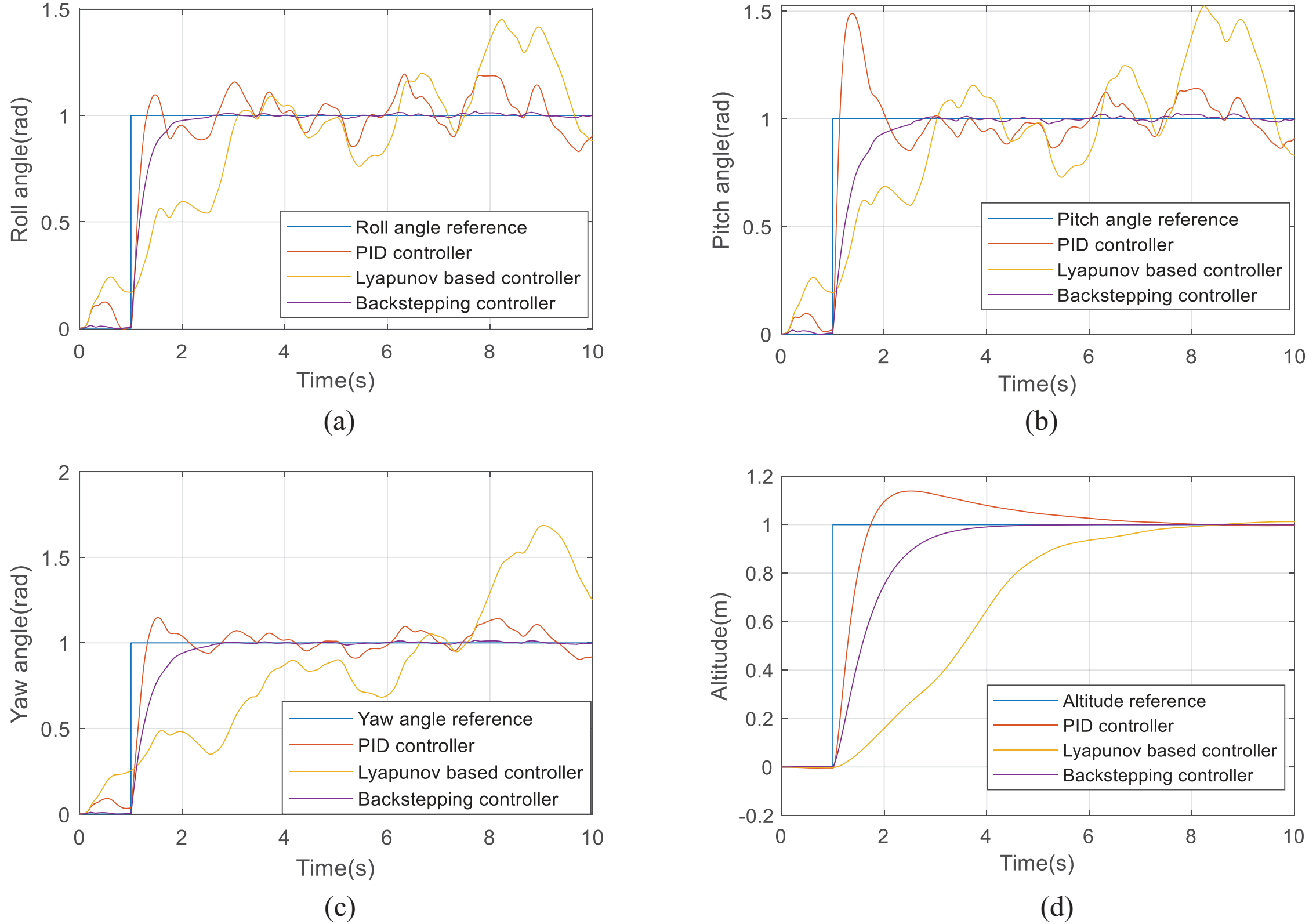

**Figure 6:** (a) Roll angle reference tracking under band-limited white Gaussian noise; (b) pitch angle reference tracking under band-limited white Gaussian noise; (c) yaw angle reference tracking under band-limited white Gaussian noise; (d) altitude reference tracking under band-limited white Gaussian noise

**Table 5:** Time response of controllers under band-limited white Gaussian noise

| Controller | Rise time (s) | Overshoot (%) | Settling time (s) |
|---|---|---|---|
| Roll angle PID | 0.19 | 15.85 | – |
| Roll angle Lyapunov-based | 1.83 | 45.08 | – |
| Roll angle backstepping | 0.57 | 0.58 | 2.13 |
| Pitch angle PID | 0.1 | 47.01 | – |
| Pitch angle Lyapunov-based | 1.73 | 67.48 | – |
| Pitch angle backstepping | 0.77 | 1.09 | 2.56 |
| Yaw angle PID | 0.25 | 10.92 | – |
| Yaw angle Lyapunov-based | 0.73 | 13.43 | – |
| Yaw angle backstepping | 0.76 | 0.7 | 2.53 |
| Altitude PID | 0.49 | 14.37 | 6.44 |
| Altitude Lyapunov-based | 4.3 | 1.53 | 7.38 |
| Altitude backstepping | 1.39 | 0.45 | 3.6 |

It is clear that the backstepping control structure is more robust to white Gaussian noise compared to PID and Lyapunov-based controllers. It has the shortest settling time. While PID and Lyapunov-based controllers show significant overshoot, the backstepping controller presents nearly no overshoot. It can be concluded that the backstepping controller can track altitude and attitude references better than PID and Lyapunov-based controllers. PID and Lyapunov-based controllers do not have a settling time for roll, pitch, and yaw angle references, as they cannot catch and stay within 2% of the final reference value.

### 5.2 *Simulations under Pink Noise*

Pink noise is background noise in electronic devices. Pink noise with a sampling time of 0.1 s was given to the system with the Colored Noise Generator block in Simulink. Simulations of controllers under pink noise are presented in Fig. 7.

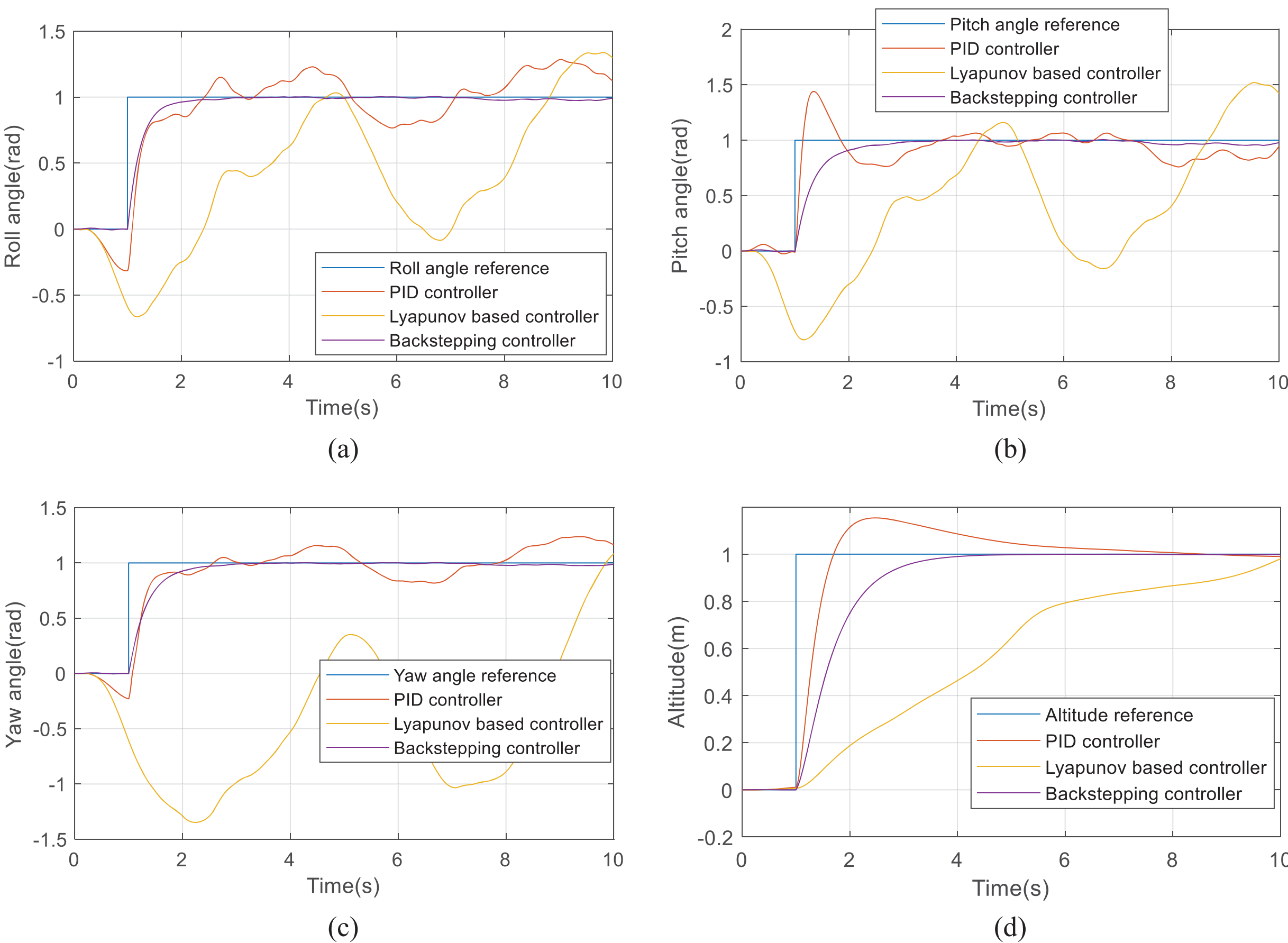

**Figure 7:** (a) Roll angle reference tracking under pink noise; (b) pitch angle reference tracking under pink noise; (c) yaw angle reference tracking under pink noise; (d) altitude reference tracking under pink noise

The comparison of rise time, overshoot, and settling time of PID, Lyapunov-based, and backstepping controllers under pink noise is shown in Table 6.

**Table 6:** Time response of controllers under pink noise

| Controller | Rise time (s) | Overshoot (%) | Settling time (s) |
|---|---|---|---|
| Roll angle PID | 1.45 | 23.28 | – |
| Roll angle Lyapunov-based | 4.69 | 34 | – |
| Roll angle backstepping | 1.39 | 0.22 | 1.83 |
| Pitch angle PID | 1.12 | 43.9 | – |
| Pitch angle Lyapunov-based | 4.29 | 52.2 | – |
| Pitch angle backstepping | 1.5 | 1.23 | 2.30 |
| Yaw angle PID | 1.40 | 8.30 | – |
| Yaw angle Lyapunov-based | 9.85 | 8.7 | – |
| Yaw angle backstepping | 1.56 | 0.34 | 2.16 |
| Altitude PID | 1.47 | 15 | 4.93 |
| Altitude Lyapunov-based | 10.16 | 0 | 9.7 |
| Altitude backstepping | 2.14 | 0 | 2.99 |

The backstepping controller has the least overshoot of all reference tracking. When following the references, it shows overshoots in the range of 0%–1%. It reaches a stable settling time value in a short time in all references.

The PID control structure represents a faster rise time compared to the other two controllers. However, it exhibits a higher overshoot compared to the backstepping controller and cannot reach a stable settling time value in reference tracking other than the altitude reference.

The Lyapunov-based controller has the longest rise time and highest overshoot compared to the other two controllers.

### *5.3 Simulations under Brown Noise*

In this section, simulations were carried out under brown noise. The equivalent of brown noise in nature is the noise made by waterfalls and rivers, heavy rainfall noise, and thunder. Brown noise with a sampling time of 0.1 s was created with the Colored Noise Generator block in the Simulink. Fig. 8 shows the simulations made under brown noise.

Table 7 represents the time response data of the controllers under brown noise.

When the data in Table 7 is examined, albeit the PID control structure has a short rise time, it shows the highest overshoot in all references. The settling time is longer than that of the backstepping controller.

Lyapunov-based controller has the longest rise time. It does not show overshoot because the rise time is very long. It also has the longest settling time. In yaw angle tracking, there is no settling time as it cannot settle within ±2% of the reference.

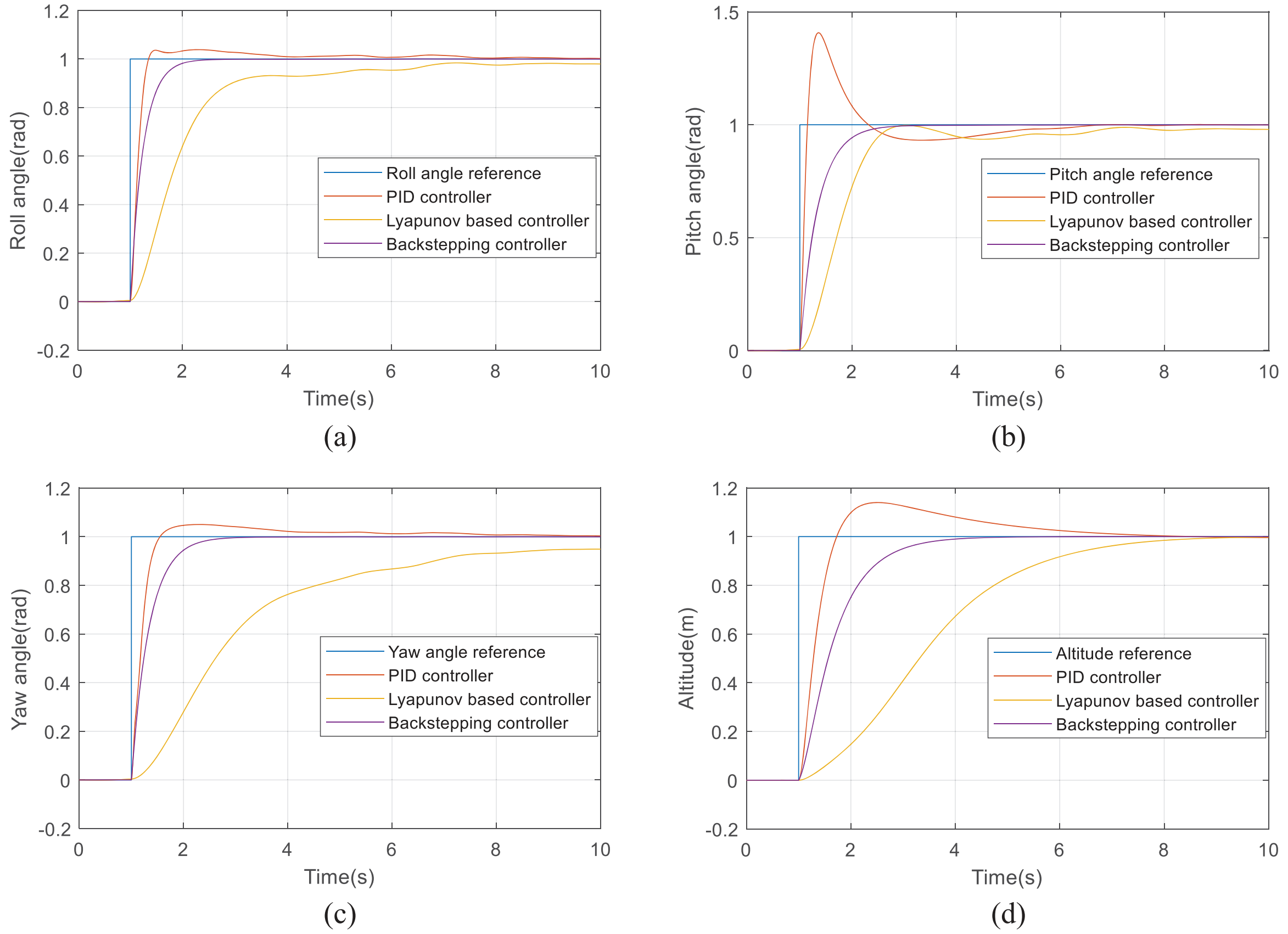

**Figure 8:** (a) Roll angle reference tracking under brown noise; (b) pitch angle reference tracking under brown noise; (c) yaw angle reference tracking under brown noise; (d) altitude reference tracking under brown noise

**Table 7:** Time response of controllers under brown noise

| Controller | Rise time (s) | Overshoot (%) | Settling time (s) |
|---|---|---|---|
| Roll angle PID | 0.24 | 3.7 | 1.33 |
| Roll angle Lyapunov-based | 1.73 | 0 | 7.01 |
| Roll angle backstepping | 0.53 | 0.39 | 1.98 |
| Pitch angle PID | 0.11 | 40.9 | 5.3 |
| Pitch angle Lyapunov-based | 1.12 | 1.53 | 6.8 |
| Pitch angle backstepping | 0.76 | 0.33 | 2.33 |
| Yaw angle PID | 0.34 | 5.1 | 4.06 |
| Yaw angle Lyapunov-based | 5.26 | 0 | – |
| Yaw angle backstepping | 0.77 | 0.41 | 2.34 |
| Altitude PID | 0.49 | 14.3 | 6.47 |
| Altitude Lyapunov-based | 3.97 | 0 | 7.70 |
| Altitude backstepping | 1.43 | 0.5 | 3.51 |

The backstepping control design presents an overshoot near 0%. Additionally, the settling time is shorter than that of other controllers in all reference tracking except for the yaw angle. The backstepping controller provides a more robust performance because it shows almost no overshoot and has a shorter settling time than other controllers.

### *5.4 Simulations under Blue Noise*

In this section, simulations were carried out under blue noise. The natural equivalent of blue noise is similar to the hissing sound made by fountains. Blue noise was produced with a sampling time of 0.1 s using the Colored Noise Generator block in Simulink. Fig. 9 shows controller simulations under blue noise.

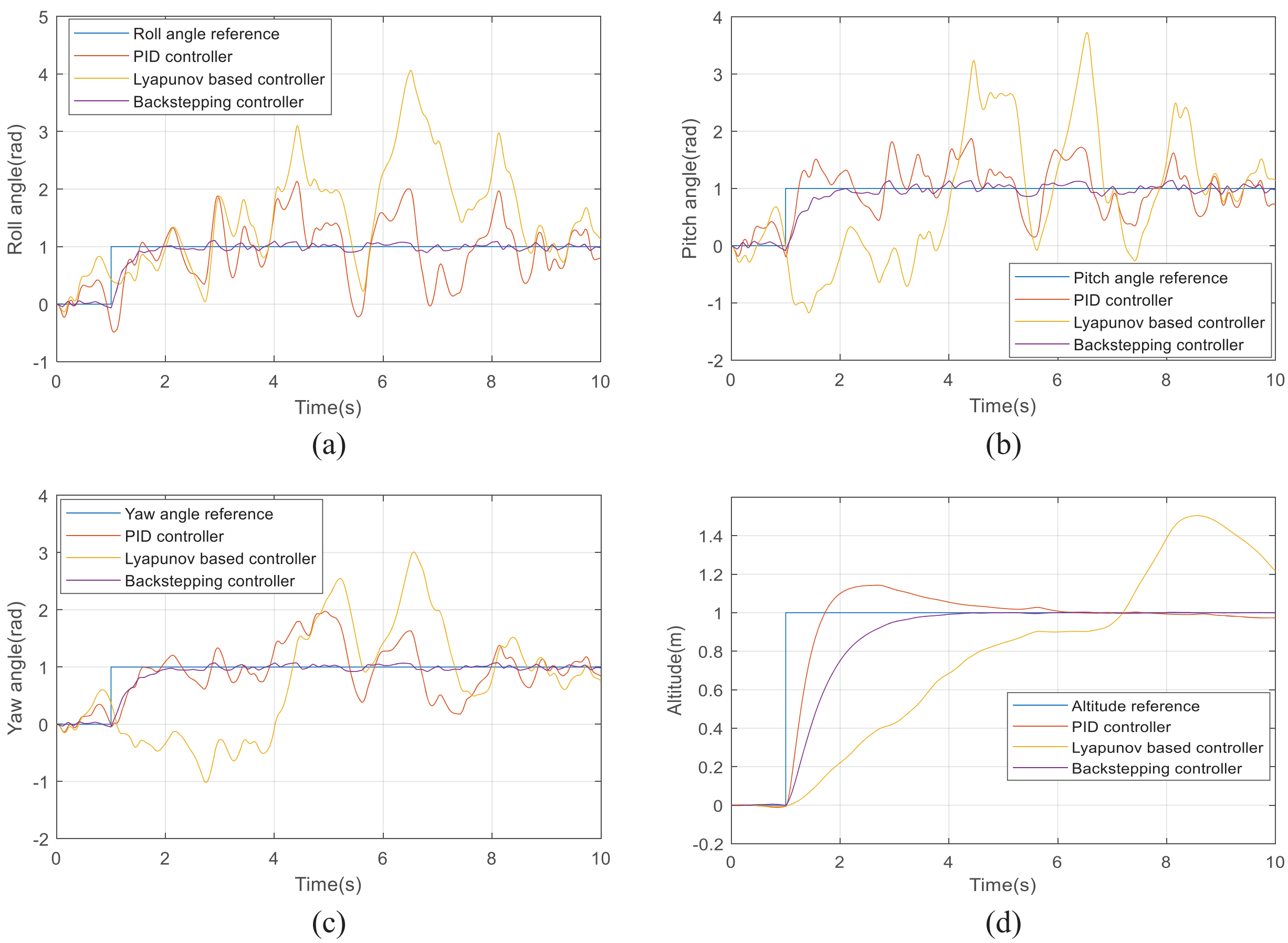

**Figure 9:** (a) Roll angle reference tracking under blue noise; (b) pitch angle reference tracking under blue noise; (c) yaw angle reference tracking under blue noise; (d) altitude reference tracking under blue noise

Table 8 represents the time response data of the controllers under blue noise.

According to Table 8, the backstepping controller has the shortest rise time for roll, pitch, and yaw angles. In addition, the backstepping controller shows the least overshoot in all reference types. The only controller that has a settling time in all references is the backstepping controller. Other controllers cannot settle within ±5% of the reference value due to noise. The PID controller has a settling time only in the altitude reference. Other references do not have data on settling time. The Lyapunov-based controller has no settling time in any reference and shows the highest overshoot. When all these results are evaluated, it turns out that the backstepping controller is the most robust of all.

**Table 8:** Time response of controllers under blue noise

| Controller | Rise time (s) | Overshoot (%) | Settling time (s) |
|---|---|---|---|
| Roll angle PID | 1.12 | 113.3 | – |
| Roll angle Lyapunov-based | 1.74 | 307.7 | – |
| Roll angle backstepping | 0.42 | 10.4 | 9.05 |
| Pitch angle PID | 0.98 | 87.2 | – |
| Pitch angle Lyapunov-based | 3.49 | 272.9 | – |
| Pitch angle backstepping | 0.77 | 13.3 | 9.07 |
| Yaw angle PID | 1.1 | 98.4 | – |
| Yaw angle Lyapunov-based | 3.88 | 203.1 | – |
| Yaw angle backstepping | 0.75 | 8.1 | 6.87 |
| Altitude PID | 0.49 | 14.2 | 4.02 |
| Altitude Lyapunov-based | 3.96 | 50.7 | – |
| Altitude backstepping | 1.42 | 0.5 | 2.57 |

### *5.5 Simulations under Purple Noise*

In this section, simulations were carried out under purple noise. The natural equivalent of purple noise is a high-pitched hissing sound. Purple noise with a sampling time of 0.1 s was generated using the Colored Noise Generator block in Simulink. Fig. 10 shows the simulations performed under purple noise.

Table 9 represents the time response data of the controllers under purple noise.

When the values in Table 9 are viewed, the backstepping control structure has the fastest rise time in all references except the altitude reference. The backstepping controller shows the lowest overshoot in all reference tracking. Additionally, it has a settling time in all references. PID and Lyapunov-based controllers have no settling time except for altitude reference because they cannot reach the $\pm 5\%$ range of the reference value. In altitude reference, the backstepping controller has a much shorter settling time than the other two controllers. In this case, it is clear that the backstepping control design is more robust than the other two controllers under purple noise.

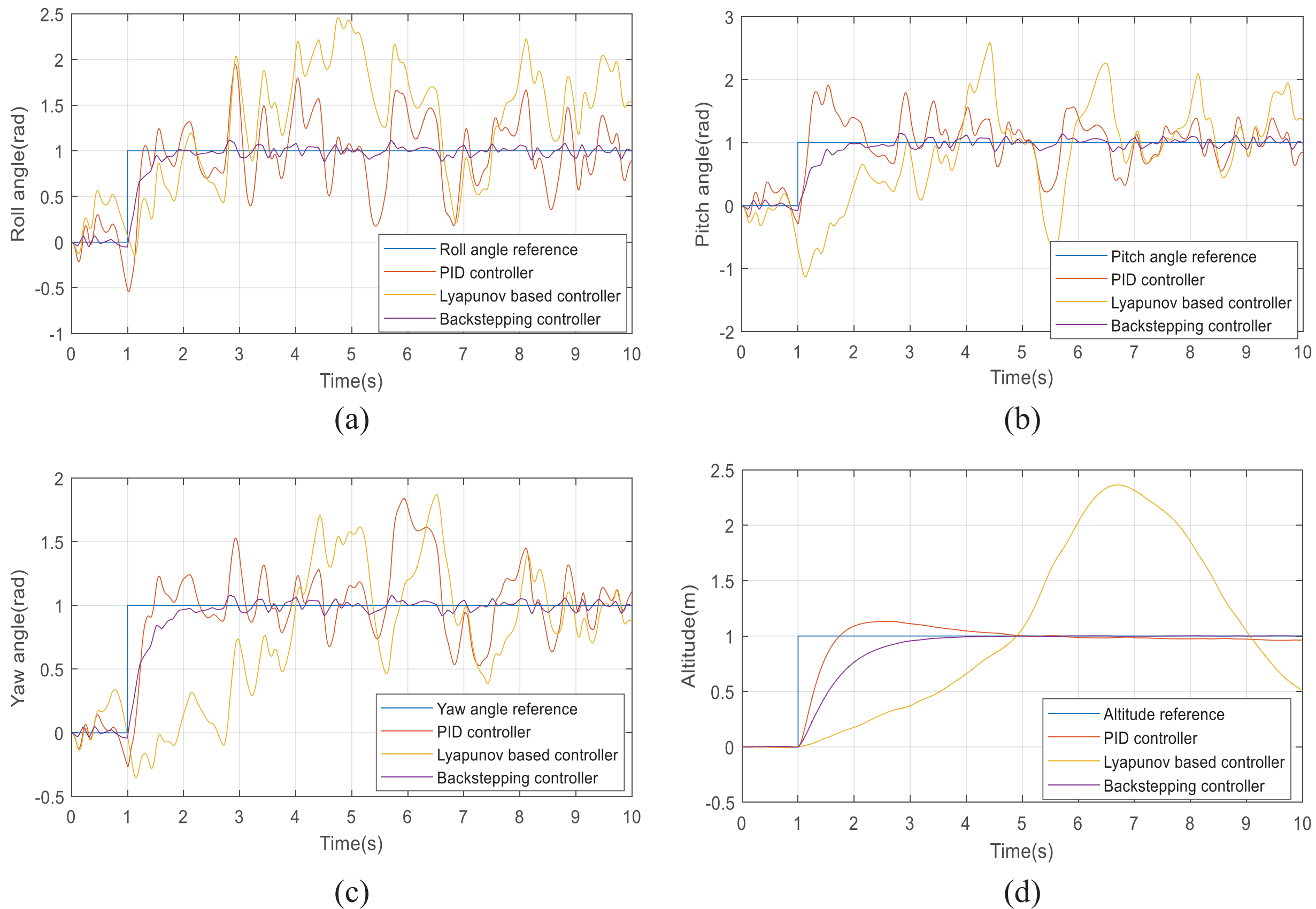

**Figure 10:** (a) Roll angle reference tracking under purple noise; (b) pitch angle reference tracking under purple noise; (c) yaw angle reference tracking under purple noise; (d) altitude reference tracking under purple noise

**Table 9:** Time response of controllers under purple noise

| Controller | Rise time (s) | Overshoot (%) | Settling time (s) |
|---|---|---|---|
| Roll angle PID | 1.04 | 95 | – |
| Roll angle Lyapunov-based | 1.77 | 145.8 | – |
| Roll angle backstepping | 0.41 | 12 | 9.85 |
| Pitch angle PID | 0.99 | 91.9 | – |
| Pitch angle Lyapunov-based | 2.22 | 158 | – |
| Pitch angle backstepping | 0.44 | 14.9 | 9.86 |
| Yaw angle PID | 0.88 | 84.3 | – |
| Yaw angle Lyapunov-based | 3.55 | 87.6 | – |
| Yaw angle backstepping | 0.77 | 8 | 6.86 |
| Altitude PID | 0.50 | 13.3 | 4.05 |

(Continued)

**Table 9 (continued)**

| Controller | Rise time (s) | Overshoot (%) | Settling time (s) |
|---|---|---|---|
| Altitude Lyapunov-based | 3.10 | 137 | 14.58 |
| Altitude backstepping | 1.42 | 0.45 | 2.93 |

## 6 Conclusion

In this research, nonlinear modeling of a quadrotor and robust backstepping controller design resistant to white Gaussian noise and colored noises were carried out. A comparative robustness analysis was performed with PID and Lyapunov-based controllers to prove the robustness of the backstepping controller. These three controllers were compared under band-limited Gaussian white noise, under pink noise, under brown noise, under blue noise, and under purple noise. Rise time, overshoot, and settling time values of controllers were acquired under all these conditions. When the results were examined, it was observed that the backstepping control structure presented the least overshoot and the shortest settling time. PID and Lyapunov-based controllers did not have settling time because they suffered excessive distortion under colored noises. PID and Lyapunov-based controllers showed high overshoot under colored noises. Additionally, the backstepping controller is the only controller with a settling time in all references. PID and Lyapunov-based controllers are less robust to noise and, therefore, do not have a settling time in many reference tracking simulations. Since the Lyapunov-based controller does not reach 90% of the given reference in almost any of the simulations, there is no rise time data. When all these results are evaluated, it is clear that the backstepping control is the most robust controller. It is the controller that shows the least overshoot and has the shortest settling time under all conditions.

## 7 Limitations and Further Study

In this study, a robust backstepping controller is designed to control the altitude and attitude of a nonlinear quadrotor model under colored noise. As seen in Fig. 1, simulations are performed under all types of noise in the color spectrum. Time response data has proven that the proposed backstepping controller is better than the classical PID and Lyapunov-based controllers. However, this study is a simulation study and has no practical application. In the future, practical applications are planned to be carried out with a real drone. In addition, it is intended to develop new methods that will enable trajectory tracking in adverse weather conditions such as fog, haze, rain, and night using a quadrotor with a camera and image processing techniques.

**Acknowledgement:** The author would like to thank the editors and reviewers for their valuable work.

**Funding Statement:** The author received no specific funding for this study.

**Availability of Data and Materials:** The author confirms that the data supporting the findings of this study are available within the article.

**Ethics Approval:** Not applicable.

**Conflicts of Interest:** The author declares no conflicts of interest to report regarding the present study.

## References

[1] F. M. Al-Qahtani, S. Elferik, and A. W. A. Saif, "Quadrotor robust fractional-order sliding mode control in unmanned aerial vehicles for eliminating external disturbances," *Aerospace*, vol. 10, no. 8, 2023, Art. no. 665. doi: 10.3390/aerospace10080665.

[2] H. Xin *et al.*, "Modeling and control of a quadrotor tail-sitter unmanned aerial vehicles," *Proc. Inst. Mech. Eng. Part I: J. Syst. Control Eng.*, vol. 236, no. 3, pp. 443–457, 2022. doi: 10.1177/09596518211050466.

[3] M. A. Fakhar and N. K. Gupta, "Performance enhancement using hybrid VTOL UAV," in *2023 Int. Conf. Sustain. Emerg. Innov. Eng. Technol. (ICSEIET)*, Ghaziabad, India, Sep. 2023, pp. 687–691. doi: 10.1109/ICSEIET58677.2023.10303571.

[4] S. H. Derrouaoui, Y. Bouzid, and M. Guiatni, "Towards a new design with generic modeling and adaptive control of a transformable quadrotor," *Aeronaut. J.*, vol. 125, no. 1294, pp. 2169–2199, 2021. doi: 10.1017/aer.2021.54.

[5] M. Karahan, M. Inal, and C. Kasnakoglu, "Fault tolerant super twisting sliding mode control of a quadrotor UAV using control allocation," *Int. J. Robot. Control Syst.*, vol. 3, no. 2, pp. 270–285, 2023. doi: 10.31763/ijrcs.v3i2.994.

[6] M. H. Sabour, P. Jafary, and S. Nematiyan, "Applications and classifications of unmanned aerial vehicles: A literature review with focus on multi-rotors," *Aeronaut. J.*, vol. 127, no. 1309, pp. 466–490, 2023. doi: 10.1017/aer.2022.75.

[7] F. Aminifar and F. Rahmatian, "Unmanned aerial vehicles in modern power systems: Technologies, use cases, outlooks, and challenges," *IEEE Electrification Mag.*, vol. 8, no. 4, pp. 107–116, 2020. doi: 10.1109/MELE.2020.3026505.

[8] B. Zhou, J. Pan, F. Gao, and S. Shen, "RAPTOR: Robust and perception-aware trajectory replanning for quadrotor fast flight," *IEEE Trans. Robot.*, vol. 37, no. 6, pp. 1992–2009, 2021. doi: 10.1109/TRO.2021.3071527.

[9] L. Jarin-Lipschitz, R. Li, T. Nguyen, V. Kumar, and N. Matni, "Robust, perception based control with quadrotors," in *2020 IEEE/RSJ Int. Conf. Intell. Robots Syst. (IROS)*, Las Vegas, NV, USA, Oct. 2020, pp. 7737–7743. doi: 10.1109/IROS45743.2020.9341507.

[10] M. A. Abitha and A. Saleem, "Adaptive PSO-tuned trajectory tracking controller for quadrotor aircraft based on Lyapunov approach," *Trans. Inst. Meas. Control*, vol. 2, no. 1, 2023. doi: 10.1177/01423312231152360.

[11] R. Kumar and S. Srivastava, "A novel dynamic recurrent functional link neural network-based identification of nonlinear systems using Lyapunov stability analysis," *Neural Comput. Appl.*, vol. 33, no. 13, pp. 7875–7892, 2021. doi: 10.1007/s00521-020-05526-x.

[12] R. Kumar, S. Srivastava, and J. R. P. Gupta, "Comparative study of neural networks for control of nonlinear dynamical systems with Lyapunov stability-based adaptive learning rates," *Arab. J. Sci. Eng.*, vol. 43, no. 6, pp. 2971–2993, 2018. doi: 10.1007/s13369-017-3034-9.

[13] L. Cheded and R. Doraiswami, "Tracking the trajectory of an object in a noisy environment with unknown statistics: A novel robust Kalman filter residue-based approach," *Trans. Inst. Meas. Contr.*, vol. 45, no. 8, pp. 1539–1557, 2023. doi: 10.1177/01423312221142119.

[14] J. Zhao, H. Zhang, and X. Li, "Active disturbance rejection switching control of quadrotor based on robust differentiator," *Syst. Sci. Control Eng.*, vol. 8, no. 1, pp. 605–617, 2020. doi: 10.1080/21642583.2020.1851805.

[15] M. E. Guerrero-Sánchez, O. Hernández-González, G. Valencia-Palomo, F. R. López-Estrada, A. E. Rodríguez-Mata and J. Garrido, "Filtered observer-based IDA-PBC control for trajectory tracking of a quadrotor," *IEEE Access*, vol. 9, pp. 114821–114835, 2021. doi: 10.1109/ACCESS.2021.3104798.

[16] M. Mahfouz, A. Taiomour, M. M. Ashry, and G. Elnashar, "PID tuning approaches for quadrotors unmanned aerial vehicles," *IOP Conf. Series: Mater. Sci. Eng.*, vol. 1172, no. 1, 2021, Art. no. 012040. doi: 10.1088/1757-899X/1172/1/012040.

[17] Z. Wang and T. Zhao, "Adaptive-based linear active disturbance rejection attitude control for quadrotor with external disturbances," *Trans. Inst. Meas. Contr.*, vol. 44, no. 2, pp. 286–298, 2022. doi: 10.1177/01423312211031781.

[18] Z. Hou, P. Lu, and Z. Tu, "Nonsingular terminal sliding mode control for a quadrotor UAV with a total rotor failure," *Aerosp. Sci. Technol.*, vol. 98, no. 3, 2020, Art. no. 105716. doi: 10.1016/j.ast.2020.105716.

[19] Q. Xu, Z. Wang, and Z. Zhen, "Information fusion estimation-based path following control of quadrotor UAVs subjected to Gaussian random disturbance," *ISA Trans.*, vol. 99, no. 3, pp. 84–94, 2020. doi: 10.1016/j.isatra.2019.10.003.

[20] R. Cen, T. Jiang, and P. Tang, "Modified Gaussian process regression based adaptive control for quadrotors," *Aerosp. Sci. Technol.*, vol. 110, no. 3, 2021, Art. no. 106483. doi: 10.1016/j.ast.2020.106483.

[21] M. Labbadi and M. Cherkaoui, "Robust adaptive global time-varying sliding-mode control for finite-time tracker design of quadrotor drone subjected to Gaussian random parametric uncertainties and disturbances," *Int. J. Control, Autom. Syst.*, vol. 19, no. 6, pp. 2213–2223, 2021. doi: 10.1007/s12555-020-0329-5.

[22] S. H. Arul and D. Manocha, "SwarmCCO: Probabilistic reactive collision avoidance for quadrotor swarms under uncertainty," *IEEE Robot. Autom. Lett.*, vol. 6, no. 2, pp. 2437–2444, 2021. doi: 10.1109/LRA.2021.3061975.

[23] O. Bouaiss, R. Mechgoug, and A. Taleb-Ahmed, "Visual soft landing of an autonomous quadrotor on a moving pad using a combined fuzzy velocity control with model predictive control," *Signal Image Video Process.*, vol. 17, no. 1, pp. 21–30, 2023. doi: 10.1007/s11760-022-02199-y.

[24] Y. Wang, J. O'Keeffe, Q. Qian, and D. Boyle, "KinoJGM: A framework for efficient and accurate quadrotor trajectory generation and tracking in dynamic environments," in *2022 Int. Conf. Robot. Autom. (ICRA)*, Philadelphia, PA, USA, May 2022, pp. 11036–11043. doi: 10.1109/ICRA46639.2022.9812352.

[25] A. Noordin, M. A. M. Basri, Z. Mohamed, and I. M. Lazim, "Position and attitude control of MAV quadrotor using super twisting sliding mode control," *AIP Conf. Proc.*, vol. 2795, no. 1, May 2023, Art. no. 40001. doi: 10.1063/5.0121378.

[26] C. D. McKinnon and A. P. Schoellig, "Estimating and reacting to forces and torques resulting from common aerodynamic disturbances acting on quadrotors," *Robot. Auton. Syst.*, vol. 123, no. 4, 2020, Art. no. 103314. doi: 10.1016/j.robot.2019.103314.

[27] G. Torrente, E. Kaufmann, P. Föhn, and D. Scaramuzza, "Data-driven MPC for quadrotors," *IEEE Robot. Autom. Lett.*, vol. 6, no. 2, pp. 3769–3776, 2021. doi: 10.1109/LRA.2021.3061307.

[28] M. Karahan and C. Kasnakoglu, "Modeling a quadrotor unmanned aerial vehicle and robustness analysis of different controller designs under parameter uncertainty and noise disturbance," *J. Control Eng. Appl. Inform.*, vol. 23, no. 4, pp. 13–24, 2021.

[29] J. Strawson, P. Cao, T. Bewley, and F. Kuester, "Rotor orientation optimization for direct 6 degree of freedom control of multirotors," in *2021 IEEE Aerosp. Conf. (50100)*, Big Sky, MT, USA, Mar. 2021, pp. 1–12. doi: 10.1109/AERO50100.2021.9438375.

[30] S. R. Nekoo, J. Á. Acosta, and A. Ollero, "Quaternion-based state-dependent differential Riccati equation for quadrotor drones: Regulation control problem in aerobatic flight," *Robotica*, vol. 40, no. 9, pp. 3120–3135, 2022. doi: 10.1017/S0263574722000091.

[31] R. P. Borase, D. K. Maghade, S. Y. Sondkar, and S. N. A. Pawar, "Review of PID control, tuning methods and applications," *Int. J. Dyn. Control*, vol. 9, no. 2, pp. 818–827, 2021. doi: 10.1007/s40435-020-00665-4.

[32] A. K. Khalaji and H. Tourajizadeh, "Nonlinear Lyapounov based control of an underwater vehicle in presence of uncertainties and obstacle," *Ocean Eng.*, vol. 198, no. 11, 2020, Art. no. 106998. doi: 10.1016/j.oceaneng.2020.106998.